\documentclass[aps,twocolumn,
superscriptaddress,
nobibnotes,
nofootinbib,
prl]{revtex4-2}

\usepackage{amsmath,amssymb,bm}
\usepackage{graphicx}
\usepackage{epstopdf}
\usepackage{latexsym}
\usepackage[normalem]{ulem} 
\usepackage[caption=false, position=top]{subfig}
\usepackage[usenames,dvipsnames]{color}
\usepackage{hyperref}
\usepackage{natbib}
\usepackage{xcolor}
\usepackage{physics}
\usepackage{multirow}
\usepackage{stackengine}
\usepackage{dsfont}
\usepackage{bbold}
\usepackage{nicefrac}
\usepackage[inkscapelatex=false]{svg}
\usepackage{enumitem}

\usepackage{framed}

\usepackage[utf8]{inputenc} 

\newcommand{\ind}[1]{_\text{#1}}
\renewcommand{\vec}[1]{\mathbf{#1}}
\newcommand{\su}{\uparrow}
\newcommand{\sdown}{\downarrow}
\newcommand{\sx}[1]{\sigma^x_{#1}}
\newcommand{\sy}[1]{\sigma^y_{#1}}
\newcommand{\sz}[1]{\sigma^z_{#1}}

\renewcommand{\a}{\vec{a}}
\newcommand{\comment}[1]{}
\renewcommand{\emph}{}
\newcommand{\errorbarstatement}{Error bars are standard deviations across multiple independent tomography datasets and neural network training runs with randomly initialized parameters.}

\def\equationautorefname~#1\null{Equation~(#1)\null}
\begin{document}

\title{Efficient quantum state tomography with convolutional neural networks}

\date{\today}

\author{Tobias Schmale}
\email{tobias.schmale@kip.uni-heidelberg.de}
\author{Moritz Reh}
\email{moritz.reh@kip.uni-heidelberg.de}
\affiliation{Kirchhoff-Institut f\"{u}r Physik, Universit\"{a}t Heidelberg, Im Neuenheimer Feld 227, 69120 Heidelberg, Germany}
\author{Martin G\"{a}rttner}
\email{martin.gaerttner@kip.uni-heidelberg.de}
\affiliation{Kirchhoff-Institut f\"{u}r Physik, Universit\"{a}t Heidelberg, Im Neuenheimer Feld 227, 69120 Heidelberg, Germany}
\affiliation{Physikalisches Institut, Universit\"at Heidelberg, Im Neuenheimer Feld 226, 69120 Heidelberg, Germany}
\affiliation{Institut f\"ur Theoretische Physik, Ruprecht-Karls-Universit\"at Heidelberg, Philosophenweg 16, 69120 Heidelberg, Germany}

\begin{abstract}
Modern day quantum simulators can prepare a wide variety of quantum states but the accurate estimation of observables from tomographic measurement data often poses a challenge. We tackle this problem by developing a quantum state tomography scheme which relies on approximating the probability distribution over the outcomes of an informationally complete measurement in a variational manifold represented by a convolutional neural network. We show an excellent representability of prototypical ground- and steady states with this ansatz using a number of variational parameters that scales polynomially in system size. This compressed representation allows us to reconstruct states with high classical fidelities outperforming standard methods such as maximum likelihood estimation. Furthermore, it achieves a reduction of the estimation error of observables by up to an order of magnitude compared to their direct estimation from experimental data.
\end{abstract}

\maketitle  

\section{Introduction}
With modern day noisy intermediate scale quantum (NISQ) \cite{preskillQuantumComputingNISQ2018} simulators outperforming each other in terms of system size and complexity on a timescale of mere months, characterizing the physically prepared states becomes exceedingly difficult. Quantum state tomography (QST) \cite{parisQuantumStateEstimation2004} describes the reconstruction of a density matrix from experimental measurement data and might be considered one of the hardest feats, exhausting not only experimental, but also numerical resource limits. This is due to the curse of dimensionality, inherent to all quantum systems, from which QST suffers in two-fold form: Not only do standard tomography schemes require an exponential amount of experimental data, but also the classical post-processing is often of intractable nature. We consider four main properties to be desirable for QST-schemes:

\begin{enumerate}[label=(\roman*), topsep=1pt,itemsep=-1ex,partopsep=0ex,parsep=1.5ex]
    \item Sub-exponential scaling in required experimental data.
    \item Sub-exponential scaling in classical post-processing.
    \item 'Observable universality', requiring that upon performing a successful tomography, any linear or non-linear quantum observable should be faithfully reconstructable, without requiring further experimental data.
    \item 'State universality' meaning that the algorithm should be indifferent to the (possibly mixed) target state that is prepared experimentally.
\end{enumerate}

Obviously, no algorithm can exist that perfectly satisfies all these conditions. Most tomography schemes that have recently been developed or applied to experimental systems, give up one or more of these conditions in order to gain w.r.t.\ the remaining. Standard maximum likelihood estimation (MLE) \cite{lvovsky_iterative_2004} scales exponentially, giving up on requirements (i) and (ii). Some Bayesian methods \cite{blume_kohout_optimal_2010} gain on (i) by giving up (ii). Many entanglement detection schemes \cite{harney_entanglement_2020, harney_mixed_2021} as well as shadow tomography \cite{Huang2020, Struchalin2020} give up on requirement (iii). A multitude of variational approaches have been developed which restrict the state space in which they seek for an optimal solution, therefore giving up on property (iv) and possibly (iii). Examples are matrix-product state tomography \cite{Cramer2010, Baumgratz2013, Lanyon2017}, which restricts its search space to weakly entangled states; compressed sensing (CS) \cite{Gross2010, schwemmerExperimentalComparisonEfficient2014, Riofrio2017}, which restricts to low-rank density matrices, and permutationally invariant (PI) tomography \cite{Toth2010, Moroder2012}, which restricts to PI states.
It is this restriction of the search space, that allows these methods to obtain target state approximations from datasets of significantly reduced sizes, which is the feature that also our work aims to exploit.

A new class of variational ansatz functions, which has been employed for QST recently are neural network quantum states (NQS) \cite{Carleo2017, Torlai2018, Torlai2020}.
The exploration of NQS is motivated by universal approximation theorems \cite{Hornik1989, Hornik1991} and the observation that many NQS can efficiently encode volume-law entanglement and thereby have higher representational power compared to most tensor-network based approaches \cite{Deng2017, huangNeuralNetworkRepresentation2021, sharir_neural_2021} as well as a favorable generalization to higher dimensions.
These emerging neural network QST (NN-QST) approaches \cite{Torlai2018, Torlai2020, Carrasquilla2019, torlai_precise_2020, Cha2020, smith_efficient_2021, melkani_eigenstate_2020, huang_investigating_2021} are starting to receive attention from the experimental communities with applications to Rydberg-, trapped-ion and optical systems \cite{Palmieri2020, Torlai2019, neugebauer_neural_2020, tiunov_experimental_2020}.
We note that neural networks have also been applied to quantum state readout tasks without relying on NQS \cite{ahmedQuantumStateTomography2021, lohani_machine_2020, quekAdaptiveQuantumState2021, lodeOptimizedObservableReadout2021}, which we, however, exclude from our definition of NN-QST, since our focus is on efficient variational methods.

A major challenge for the successful application of NN-QST schemes is the choice of the variational ansatz, i.e. the network architecture, and the understanding of its intrinsic limitations. Recently, it was shown that convolutional neural networks (CNNs) are capable of efficiently encoding volume-law entanglement \cite{Levine2019} for pure states, motivating us to explore this architecture also for NN-QST tasks operating on mixed states. 
We use CNNs to learn a state's probabilistic representation from measured data, bridging the gap between quantum theory and neural networks by employing Positive Operator Valued Measures (POVMs). The resulting QST scheme scales subexponentially in the system size as it contains no exponentially large state representations and makes no assumption on the target state's purity, generalizing previous approaches \cite{Torlai2018, Torlai2020}. 
While this scheme has been demonstrated for specific classes of states and network architectures \cite{Carrasquilla2019}, applications to experimental systems have not exceeded the few qubit regime \cite{neugebauer_neural_2020, Cha2020}. The main reason for this are poorly understood performance advantages. Making further progress crucially requires the strengths and limitations of neural network based tomography schemes to be evaluated in comparison to standard tomography methods.
Here we perform quantitative comparisons between our NN-QST scheme and standard techniques like MLE.
For a broad range of typical experimental scenarios, we see improvements compared to MLE for small datasets, i.e. few measurement samples, as well as a noise reduction for the estimation of local observables on larger systems, thus decreasing the necessary amount of experimental samples at a given error threshold.

\makeatletter
\def\convertto#1#2{\strip@pt\dimexpr #2*65536/\number\dimexpr 1#1}
\makeatother

\begin{figure}
    \includegraphics{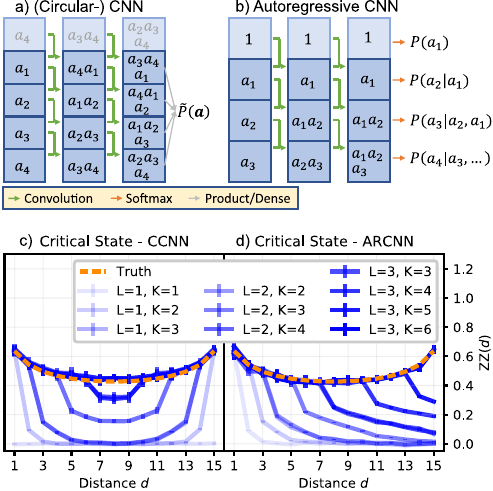}
    \caption{\textbf{CNN architectures and their expressivity.} a), b): Schematics of the CNN architectures. Entries of cells should be understood as a dependency, not as literal values, e.g.\ bottom right cell of standard CNN \textit{depends} on $a_2, a_3$ and $a_4$. Lighter cells are padding cells. The green arrows indicate the dot-product of the originating cells with a kernel. See the methods section for details. a) Standard CNN, here with circular boundary (CCNN), b) Autoregressive CNN (ARCNN). c), d): $\langle \sz{i}\sz{j}\rangle$ correlator as a function of distance $d=|i-j|$, for varying network depth $L$ and kernel size $K$, shown for 1D transverse-field Ising model (TFIM) ground states ($N=16$, $J/B=1$). \errorbarstatement \convertto{cm}{\the\textwidth}}
    \label{fig:architecture}
\end{figure}

\section{Results}
\subsection{POVM formalism}  For efficient NN-QST a suitable variational ansatz is crucial. Here we adopt an approach pioneered in \cite{Carrasquilla2019}, where the quantum state is encoded by a probability distribution over the outcomes of experimental measurements. This probability distribution, in turn, is approximated by a neural network. This has the advantage that it allows for a tomography scheme that is directly compatible with experimentally measured data and allows for applying standard probabilistic machine learning models operating on real numbers.
In the following we briefly summarize the employed learning rule and the POVM formalism, and direct to \cite{Carrasquilla2019}, as well as \cite{Carrasquilla2019a, neugebauer_neural_2020, reh2021timedependent} for further details on NQS in combination with POVMs.

In the probabilistic formulation of quantum mechanics, the state of a system is not represented by its density matrix $\rho$, but by the Born-rule probabilities 
\begin{equation}
\label{eq:Pa}
    P(\vec{a})=\text{Tr}[\rho M_\vec{a}]
\end{equation}
of an informationally complete (IC) POVM, consisting of measurement operators $M_\a$, where $\mathbf{a}$ labels possible measurement outcomes.
`Informationally complete' means that the $M_\a$ form a complete basis for the set of hermitian operators, implying that any density matrix or observable can be expanded using the POVM operators. Therefore, knowing $P(\vec{a})$ is equivalent to knowing $\rho$ and in principle $\rho$ can be inferred from $P(\vec{a})$ by inverting Eq.~\eqref{eq:Pa} to $\rho = P(\a) T_{\a \a'}^{-1} M_{\a'}$ using the overlap matrix $T_{\a \a'} = \text{Tr}[M_\a M_{\a'}]$ and by summing over repeated indices. 
Any observable may be computed by sampling from the POVM distribution using
\begin{equation}
\text{Tr}[\rho O] = \sum_\vec{a} P(\vec{a}) O_\a = \langle O_\a \rangle_{\a \sim P(\a)},
\end{equation}
where $O_\a = \text{Tr}[M_{\a'}O]T^{-1}_{\a\a'}$ is the POVM representation of an observable $O$.
We consider systems of $N$ qubits where each qubit is read out individually and thus the POVM elements are product operators $M_\vec{a}=M_{a_1}\otimes\ldots\otimes M_{a_N}$.
With such a factorized POVM, the overlap matrix $T$ also factorizes, which allows the computation of $O_\a$ to be efficient, given that $O$ is a local observable, or can be expressed as a sum of few (polynomially many in N), possibly non-local, Pauli strings \cite{Carrasquilla2019}.

The experimentally most convenient IC measurement scheme consists of single qubit Pauli measurements. By randomly selecting the $x$, $y$ or $z$ basis for measuring each qubit in each experimental run, one effectively measures a POVM with 6 possible outcomes per qubit (3 bases with 2 outcomes each). Unfortunately, this results in an overcomplete POVM, with an overlap matrix $T$ that is not invertible. This can easily be remedied by grouping three out of the six outcomes into one, resulting in four POVM operators such as e.g. $$\{M_{0/1/2} = \frac{1}{3}\ketbra{\su_{x/y/z}}, M_3 = \mathbb{1} - M_0-M_1-M_2\}.$$
Here $M_3$ now incorporates the 3 $\ket{\sdown}$ outcomes for all axes $x$, $y$ and $z$. This POVM is typically referred to as the Pauli-4 POVM \cite{Carrasquilla2019}.
In this setting, a measurement outcome $\vec{a} = a_1a_2...a_N$ is a string of $N$ single qubit outcomes for an $N$ qubit system, over which the POVM distribution $P(\vec{a}) = P(a_1, a_2, ..., a_N)$ is defined. A dataset of size $N_s$ is a set of multiple such outcomes $D = \left\{\vec{a}_1, \vec{a}_2, ..., \vec{a}_{N_s}\right\}$.

We use a neural network as a non-linear function that returns a single POVM probability $P(\vec{a})$ given a POVM outcome $\vec{a}$ as input. We can thus write our neural network ansatz for the POVM distribution as $P^\vec{\theta}\ind{NN}(\a)$, where $\vec{\theta}$ denotes the tuple of variational parameters describing the ansatz.
Fitting the variational parameters of the network builds on standard machine learning methods: We use the established \textsc{Adam} optimizer \cite{kingma_adam_2017} to find the parameters $\vec{\theta}$ that have the highest likelihood of reproducing the data. This is equivalent to minimizing the cross-entropy between the dataset distribution and the NN distribution.
The \textsc{Adam} optimizer gave considerably better results than less advanced optimizers like the pure gradient-descent optimizer or a momentum optimizer.

We finally point out, that the positivity condition on the density matrix is not known to easily translate into the POVM formalism, without introducing exponential scaling \cite{Fuchs2013}. Therefore, one cannot guarantee that all sampled observables obey physical constraints (such as $\text{Tr}[\rho^2] < 1$). 
We have observed violations of such constraints very rarely, hence we argue that the lack of the positivity condition is not a limiting factor of this approach.
While RBM purifications \cite{Torlai2018a} are one means of achieving a positive neural network density matrix,  these can currently not make use of other, more flexible network architectures such as CNNs.

\subsection{CNN architectures} NN-QST in conjunction with POVMs has been successfully applied to recurrent neural networks (RNNs) \cite{Carrasquilla2019} as well as attention based models \cite{Cha2020}. We expand on this knowledge, by employing two distinctly different versions of CNNs, motivated by recent theoretical developments, which frame CNNs as a generalization to matrix product states \cite{Levine2019, sharir_neural_2021}. 

The first architecture we consider is the 'standard' CNN in Fig.~\ref{fig:architecture} a), which is fed a one-hot encoded vector of single-site outcomes, performs repeated convolutions on it and returns the corresponding (unnormalized) POVM probability $\tilde{P}\ind{NN}(\a)$. 
Motivations for using this architecture are its established representational power when operating on pure states \cite{Levine2019, sharir_neural_2021}, the ability to represent 1D and 2D states, as well as the ability to encode symmetries such as translation invariance without increasing the computational complexity. Unfortunately, one has to resort to Markov Chains in order to draw samples from the network, which can lead to unintended correlations between generated samples \cite{liuMonteCarloStrategies2004}.

The second architecture is a 1D autoregressive CNN (ARCNN) \cite{Sharir2019, lin_scaling_2021}, illustrated in Fig.~\ref{fig:architecture} b). This architecture is very similar to the standard CNN, but makes use of the autoregressive property, which states that a probability of multiple variables can be partitioned into a product of conditionals, which are returned by the final layer of the ARCNN: $P(\textbf{a})=P(a_1) P(a_2|a_1)\ldots P(a_N|a_1..a_{N-1})$. 
Other network architectures that also make use of this property include recurrent neural networks \cite{Carrasquilla2019} and Transformers \cite{Cha2020}.
The autoregressive property of the ARCNN allows probabilities to be exactly normalized, and for samples to be drawn exactly, circumventing the need for Markov Chains.
More details on both architectures can be found in the methods section.

\subsection{Expressivity of CNNs} For both the CNN with product-output layer (see methods) and the ARCNN, one can write down a maximum (physical) distance $d\ind{max}$, beyond which correlations are typically not captured correctly anymore:
\begin{align}
    d\ind{max}^\text{CNN} &= (K-1)L,\label{eq:correlation_limits}\\ 
    d\ind{max}^\text{ARCNN} &= (K-1)L + 1 \,,
\end{align}
where $L$ is the network depth and $K$ the size of the convolution kernels. 
Both can be understood intuitively from the schematics in Fig.~\ref{fig:architecture} a) and b): With each layer (i.e. $L$ times), the dependence of any single site outcome gets propagated by a distance of one less than the kernel size (i.e. $K-1$). For the autoregressive CNN, an input of $a_i$ results in the probability for $a_{i+1}$, explaining the fact, that correlations reach one site further as compared to the CNN.
These cutoff points for correlations can be seen in Fig.~\ref{fig:architecture} c) and d).

For the following results, we are mainly interested in systems with long-range correlations, hence we use architectures with $d\ind{max} = l$, where $l$ is the (side-) length of (2D) 1D systems. This gives a very compact ansatz, with the number of variational parameters scaling only cubically in the system size.
For the benchmark cases studied in this work, we did not find a benefit in increasing $d\ind{max}$ beyond this point, as this increases the number of network parameters, artificially enlarging the search space and increasing the risk of overfitting, resulting in worse generalization.
In the following it will be crucial to quantify how well the networks generalize, i.e. how many samples one needs to learn a state well, or equivalently, what errors one can expect for given dataset sizes.

\begin{figure}
    \centering
    \includegraphics{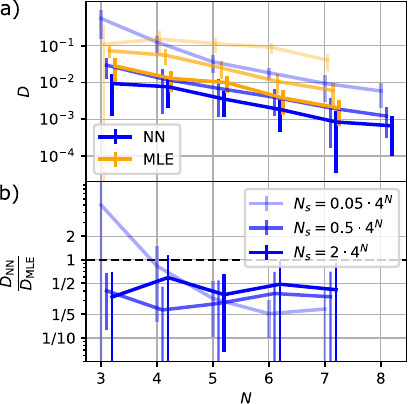}
    \caption{\textbf{Tomography of 1D Ising groundstates: NN vs MLE.} a) Residual classical infidelity of CCNN and MLE on 1D Ising ground state with periodic boundary conditions, $J/B = 1$. Shown for different dataset sizes $N_s$. b) Residual network infidelity normalised to residual MLE infidelity. \errorbarstatement}
    \label{fig:ising1d}
\end{figure}

\subsection{NN-QST benchmarks} We benchmark the tomography scheme by comparing it to two conventional schemes: MLE, i.e. parametrizing the full density matrix and maximizing the likelihood of this reproducing the dataset, as it presents the go-to choice for many small scale qubit systems and is commonly used in experiments \cite{lanyon_measurement_2013, lvovsky_iterative_2004}, as well as direct estimation of observables from the dataset.
To carry out this comparison we generate synthetic measurement data sets. We compute a target density matrix exactly and compute its POVM distribution, from which we draw samples ($1k-100k$ for 16 qubit systems). We then use these samples to train the network, resulting in the optimal variational representation $P\ind{NN}(\a)$. 
For small systems, i.e. those where it is feasible, we perform MLE to obtain an estimate for the density matrix following \cite{lvovsky_iterative_2004}, of which we compute the POVM distribution $P\ind{MLE}(\a)$. We then ask, which of these two estimates is closer to the ground truth target distribution, as measured by the classical infidelity
\begin{equation}
    D\ind{NN/MLE} = 1 - \sum_\a \sqrt{P\ind{NN/MLE}(\a)P\ind{Truth}(\a)}.
\end{equation}
This allows us to quantify, which of the two estimates is better, by using the quotient $D\ind{NN}/D\ind{MLE}$. If it is less than one, the network gives the better estimate for the target state compared to MLE, and vice versa.

For systems where MLE is infeasible, we instead consider the root mean square (RMS) error of observables
\begin{equation}
    \text{RMS}\ind{NN/Data} = \sqrt{\left\langle(O\ind{NN/Data} - O\ind{Truth})^2\right\rangle}\,,
\end{equation}
inspired by \cite{torlai_precise_2020}. These can either be computed from the training dataset itself, or from the network-encoded distribution, from which we draw 500k samples for 16 qubit systems. 
In this situation the NN acts in a way of replacing the measured dataset with a larger, network-generated one, aiming to decrease statistical measurement noise.
Here we can ask, if the NN gives an advantage for the estimation of observables, by looking at the quotient $\text{RMS}\ind{NN}/\text{RMS}\ind{Data}$.

\begin{figure}
    \centering
    \includegraphics{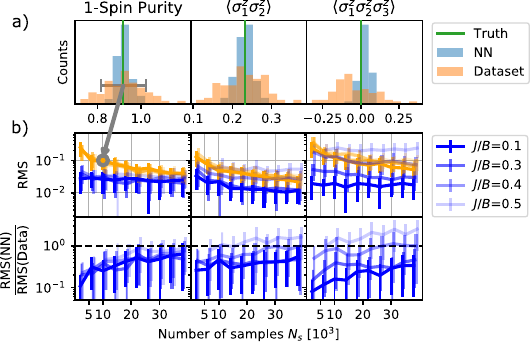}
    \caption{\textbf{Tomography of 2D Ising groundstates: NN vs plain data.} Comparing the performance of the CCNN against the plain dataset in estimating local observables on a $4\times 4$ Ising lattice with periodic boundaries. a) Histogram of observable estimates from independent datasets and network initializations, with $J/B = 0.3$ and $N_s = 10k$. b) RMS errors of observables for varying coupling strengths and dataset sizes, as well as comparisons of NN and plain dataset in terms of the quotient of their respective RMS errors. \errorbarstatement }
    \label{fig:ising2d}
\end{figure}

\subsubsection{Transverse field Ising model} We start by benchmarking this method on ground states of a translation-invariant TFIM
\begin{equation}
    H = -J \sum_{\langle i, j \rangle} \sz{i} \sz{j} - B \sum_i \sx{i}\label{eq:Hising}
\end{equation}
with coupling strength $J > 0$ using the standard CNN. This serves as a proof of concept in an idealized scenario, as the translation invariance is directly encoded in the neural network. Since the comparison of a symmetrized network to an unsymmetrized reference is somewhat unfair, we refrain from enforcing any symmetries for the later examples.

Figure~\ref{fig:ising1d} shows the method being applied to small, i.e. MLE-suitable 1D states at the Ising-critical point. Our method achieves a reduction of infidelity by a factor 2-5, depending on system and dataset size. 
The figure shows one main trend: the network advantage shrinks for increased dataset size. This is also an expected result, as MLE has to outperform any variational approach in the limit of infinite dataset size since the network is an approximation while MLE uses a full parameterization of the state. 
We generally see this behaviour for all studied systems.

In Fig.~\ref{fig:ising2d} we show the method being applied to a $4\times 4$ lattice, which is not feasible for MLE anymore. The histograms show how enhancing the dataset using the NN can lead to a reduced variance of observable estimates. For the network advantage, we see two trends: An increased advantage for small datasets, as well as an improved performance for smaller coupling strengths. The latter results in states closer to product states, which are easier for the network to learn. If $J/B$ becomes too large, training the network becomes unreliable on such small datasets and the advantage disappears.

\begin{figure}
    \centering
    \includegraphics{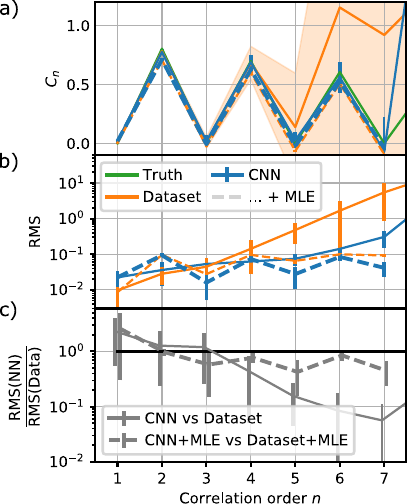}
    \caption{\textbf{Tomography of noisy long-range interacting groundstates: NN vs plain data.} Evaluating the ARCNN by evaluating observables from Eq.~\eqref{eq:corrPower} and comparing to those obtained from the plain dataset, on a length 16 ion chain with 3\% dephasing noise, $J/B = 0.6$ and open boundaries, with a dataset size of $N_s = 10$k. Dashed lines show results of applying local MLE directly to dataset (orange) and to a neural network enhanced dataset (blue). a) Observables according to Eq.~\eqref{eq:corrPower}, b) residual RMS error, c) residual network RMS normalized to dataset RMS. \errorbarstatement}
    \label{fig:ion_chain}
\end{figure}

\subsubsection{Noisy long-range interacting ion chain} For a more experimentally motivated \cite{tan_observation_2021} example, we look at ground states $\ket{\psi_0}$ of a 16-site, long-range interacting ion chain Hamiltonian
\begin{equation}
    H = -\sum_{i, j} \frac{J}{|i-j|^{1.1}} \sz{i} \sz{j} - B \sum_i \sx{i}
\end{equation}
with $J > 0$, open boundary conditions and small ($3\%$) added dephasing noise. The target state is thus $\rho\ind{Target} = 0.97 \ket{\psi_0}\bra{\psi_0} + \frac{0.03}{2^{16}}\mathbb{1}$. As this is naturally a 1D system, we use the ARCNN. Here we study the NN advantage for correlation functions of increasing order. This is interesting, as higher order correlators are typically harder to estimate from samples, as the variance of the POVM-observable scales exponentially in correlation order.
Thus, higher moments require a better approximation of the state, providing a sensitive benchmark for the quality of the state representation.
Specifically, we look at powers $n$ of the Pauli-Z operator
\begin{equation}
C_n := \frac{1}{16-n+1}\sum_{i=1}^{16-n+1}\left\langle \sz{i} \sz{i+1}...\sz{i+n-1}\right\rangle. \label{eq:corrPower}   
\end{equation}
Local observables, like the terms in Eq.~\ref{eq:corrPower}, only depend on the reduced density matrix of the subsystem they act on. Thus, these observables can in principle be estimated by performing MLE on this subsystem only. We show this 'local MLE' applied to the bare dataset and to the NN-enhanced dataset in Fig.~\ref{fig:ion_chain} in addition to the previously employed benchmarks. Using the NN-generated dataset, we see a reduction in RMS to a degree, that allows sampling for correlators of three orders higher, than what is possible with the bare dataset. When applying the local MLE to both bare and NN-generated dataset, this advantage is reduced significantly, but does not disappear. However, we emphasize that the ARCNN is on par with MLE, at a greatly reduced computational complexity. We note that the computational cost depends on the specific implementation and that there exist more efficient MLE algorithms than the one used here (eg. \cite{shangSuperfastMaximumlikelihoodReconstruction2017}). However, the difference in scaling behaviour between MLE and NN-QST persists (cf. Methods section).
Notice also the peak in RMS at a correlation order of 2 for MLE, leading to an increased RMS compared to plain sampling.
We find this to be systematic, which is why we refrain from doing this comparison to local MLE in the remainder of this work, where second order correlation functions are of interest.

Close examination of the datapoint for $C_6$ in Fig.~\ref{fig:ion_chain} reveals effects of the lack of a positivity constraint on the density matrix for direct estimation from samples and POVM based NN-QST. Here the dataset estimate lies outside the physical range $-1 \leq C_n \leq 1$, but the network is able to correct this due to the reduced statistical error. Only for $C_8$ (not shown) the network cannot cure the unphysical feature present in the dataset. Testing that observables only take physical values may serve as a useful sanity check for whether the network generalizes towards a physical state.

\begin{figure}
    \centering
    \includegraphics{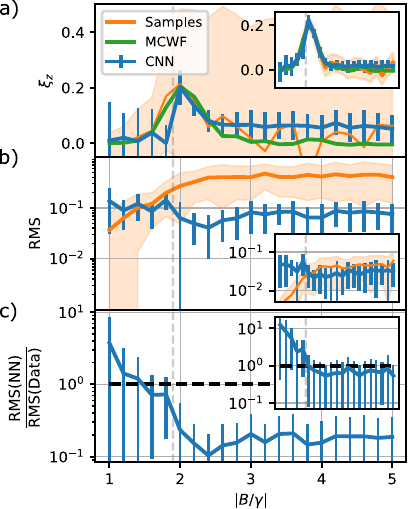}
    \caption{\textbf{Tomography of steady states: NN vs plain data.} Dissipative phase transition of a $4\times4$ TFIM with spontaneous emission at $J=-1.25\gamma$ with open boundaries. $N_s = 1$k (insets: 100k). Gray vertical line: Switch $\uparrow$ and $\downarrow$ in POVM for CNN. a) Observables according to Eq.~\eqref{eq:corrLen} but summed only over system diagonal, b) residual RMS error, c) residual network RMS normalized to dataset RMS. \errorbarstatement} 
    \label{fig:steadyStates}
\end{figure}

\subsubsection{Steady states of a driven dissipative 2D-system} As a final system, we consider steady states of a $4\times4$ TFIM with spontaneous decay, motivated by ongoing research into phase diagrams of open quantum systems \cite{jin_phase_2018}, with potential applications to Rydberg systems. We switch back to the standard CNN with dense output layer, since the system is two-dimensional. However, now no symmetries are enforced in the network.
We use the Monte-Carlo wave function approach \cite{Plenio1998}, to simulate the dynamics under the Lindblad master equation
\begin{equation}
\dot{\rho} = -i [H, \rho] + \gamma \sum_j \left(L_j\rho L_j^\dagger - \frac12 \left\{L_j^\dagger L_j, \rho\right\}\right)
\end{equation}
with the Hamiltonian from Eq.~\eqref{eq:Hising} with antiferromagnetic coupling $J < 0$, $B<0$ and
\begin{align}
L_j &= \sigma^-_j=\frac{1}{2}\left(\sx{j} - i \sy{j}\right),
\end{align}
until the steady state is reached. We consider the density matrix that is obtained using 1000 pure-state trajectories as our exact target.
This system undergoes a dissipative phase transition \cite{jin_phase_2018} which is visible as a peak in the correlation length
\begin{equation}
\xi_z^2= \sum_{i, j} |\vec{r}_i-\vec{r}_j|^2\left(\langle \sz{i}\sz{j}\rangle - \langle \sz{i}\rangle\langle \sz{j}\rangle\right) \,. \label{eq:corrLen}
\end{equation}

We show this phase transition as obtained by computing the correlation length on the 1D diagonal of the 2D lattice, once directly from the training data, as well as from a NN-enhanced dataset in Fig.~\ref{fig:steadyStates}. The network is able to capture the phase transition, as a peak in the correlation length at $|B/\gamma| \approx 2$ is clearly visible. 
At the dashed gray line, we exchange $\su$ and $\sdown$ in the POVM that the CNN uses (i.e. $M_3$ now groups all the $\ket{\su}$ outcomes instead of the $\ket{\sdown}$ outcomes), ensuring that the target state does not contain exact zeros in its POVM distribution.
For small $|B|$, the steady state tends towards an eigenstate of the observable in question. Thus the variance of sampling this observable is significantly reduced and the network has no advantage here.
For the limit of large $|B|$ we see a huge variance in the sampled correlation length, which the CNN trades for a small bias, i.e. systematic error. The overall effect is that the CNN bias and CNN variance lead to a significantly smaller RMS error as compared to the bare dataset. Notice that this bias also shrinks with increasing dataset size (Fig.~\ref{fig:steadyStates}, insets).

Depending on the observable of interest, the bias can have a more severe effect than depicted. When computing the correlation length over the 1D diagonal, as in Fig.~\ref{fig:steadyStates}, the corresponding sum in Eq.~\eqref{eq:corrLen} is a weighted average of ${4\choose2} = 6$ connected correlators of the form $\langle \sz{i}\sz{j}\rangle - \langle \sz{i}\rangle\langle \sz{j}\rangle$. After sampling, one may consider each of these connected correlators as a random variable with a variance and a bias. For the plain dataset, this bias is of course zero. However, when evaluating Eq.~\eqref{eq:corrLen} over the entire lattice, the sum contains ${16\choose2} = 120$ terms, with roughly similar variance and bias.  For the sampled case, by simple addition of probability distributions, the variance of the latter observable is thus reduced by a factor of $\sqrt{120/6} \approx 4.5$ compared to the former one. Due to the bias, the network is not able to make use of this self averaging effect, resulting in a significantly reduced advantage. We show this full system correlation length in the methods section.
Thus the network advantage is the largest, if one is interested in the expectation values of individual correlators, and might be smaller if large sums over many similarly distributed correlators are involved as the bias inherent to the variational approach becomes more statistically significant for the averaged results.

\section{Discussion} Motivated by the proven superior representation capabilities of CNNs \cite{Levine2019}, we explored the application of two different CNN architectures to NN-QST tasks, i.e.\ the reconstruction of pure and mixed quantum states. We especially found the autoregressive CNN to be extremely versatile, due to its great expressivity, exact sampling and stability during training. For a broad range of experimentally relevant scenarios, including pure and dephased ground states, as well as steady states, we presented quantitative comparisons to traditional schemes. This showed a significant advantage over MLE or, for system sizes inaccessible to MLE, direct sampling of local observables from an IC-POVM dataset. Although we demonstrated the described properties on synthetic data, the method may be readily applied to real quantum simulation experiments. 
Once the quantum state has been successfully learned from experimental data, any linear or non-linear observable can be extracted from the resulting NQS representation.

Like any variational approach, which benefits from restricting the state space to a representable subspace, this scheme is subject to potential bias. The effect of such bias intimately depends on the estimated observable, the approximated state, as well as the sample size, as discussed above. This makes it necessary to a priori validate the method to ensure generalization of a given model to the situation at hand, as we have done in this work for a range of experimentally motivated cases.

We emphasize that the proposed method can harness the strengths of any variational function approximator, thus directly profiting from the rapid development of ever more expressive architectures in the machine learning community. 
New network architectures can only enhance the state space that is covered by NN-QST, motivating further research in this area.
For future projects it would be interesting to compare many of the NN-QST schemes, to methods like shadow tomography \cite{Huang2020, Struchalin2020}, which explicitly claim superiority over NN-QST.

\section{Methods}

\subsection{Details on network architectures} 
To better explain the obtained parameter counts for the neural networks, we briefly explain the convolution operation in more detail, which the CNNs perform iteratively (here shown for the 1D case). 
\begin{align}
\left(\text{conv}(\mathbf{x};\ \mathbf{k}, b)\right)_i = f\left(b + \sum_{j=0}^{|\mathbf{k}|-1} k_j x_{j+i}\right), \notag\\ \text{for } i \in \{0, ..., |\mathbf{x}| - |\mathbf{k}|\} \label{eq:convolution}
\end{align}
of the input data $\mathbf{x}$ with a so-called kernel $\mathbf{k}$ denotes a single convolution. Here the length of a vector $\mathbf{v}$ is denoted by $|\mathbf{v}|$, $b$ is a bias and $f$ is again a non-linear activation function acting element-wise on the results of the convolution. This can be thought of as taking dot-products of the kernel and translated section of the input vector. For a 2D CNN, $\mathbf{k}$ would be a matrix, and Eq.~\eqref{eq:convolution} would compute dot products between this matrix and translated submatrices of the then two-dimensional input $\mathbf{x}$.

Multiple kernels are used per layer $l$ resulting in multiple intermediate representations $\mathbf{x}^{l, m}$. The latter is then computed via 
\begin{align}
\mathbf{x}^{l+1, m} = \sum_n \text{conv}(\mathbf{x}^{l, n};\ \mathbf{k}^{l, m, n}, b^{l, m}).\label{eq:cnn}
\end{align}
This structure on kernel-level is \emph{not} explicitly depicted in Fig.~\ref{fig:architecture}. The axis indexed by $m$ here is often called the \emph{feature} dimension.

Keeping the number of features per layer $f=N$ constant, as done throughout this work and using Eq. \eqref{eq:correlation_limits}, the $\mathbf{k}^{l, m, n}$ tensor has $K\cdot L\cdot f \cdot f = \mathcal{O}(N) \cdot N \cdot N = \mathcal{O}(N^3)$ parameters.

We employ two distinctly different network architectures. The first is the 'standard' CNN in Fig.~\ref{fig:architecture} a), which is fed a one-hot encoded vector of single-site outcomes (leading to an input shape of (Batch-Size, $N$, 4)) and performs $L$ convolutions with kernels of size $K$. Boundary conditions are either open or periodic, depending on the symmetries of the target state. In Eq.~\eqref{eq:convolution} a periodic boundary condition implies that $x_{i+j}$ wraps around to $x_0$ when $i+j$ reaches the length of $x$. A final layer turns the network output into a single scalar. Two common options are a dense layer, or a product layer, the latter resulting in network outputs of the form $\tilde{P}\ind{NN}(a) = e^{\sum \text{last layer}}$. This results in an \emph{un-}normalized probability distribution. Therefore, before each training step, a Monte Carlo estimate of the normalization constant has to be performed. For normalization, we generate as many uniform POVM samples as there are samples within each training batch. For translation invariant states, we use the product output layer and the dense output layer otherwise.

The second architecture is a 1D autoregressive CNN (ARCNN) \cite{Sharir2019, lin_scaling_2021} in Fig.~\ref{fig:architecture} b). This architecture makes use of the autoregressive property, which states that a probability of multiple variables can be partitioned into a product of conditionals: $P(\textbf{a})=P(a_1) P(a_2|a_1)\ldots P(a_N|a_1..a_{N-1})$, see \cite{lin_scaling_2021} for more details on the autoregressive structure.
The autoregressive CNN only differs from the standard CNN in the boundary conditions and in the fact that the physical dimension is shifted by one site, so that the last site is not used as an input. The outputs are fed into a softmax layer, which results in four conditional probabilities to be returned from each site. The exactly normalized probabilities may be computed by passing a POVM outcome through the network and selecting the conditional probability according to the input at each site. Exact samples can be drawn from this distribution, by passing zeros through the network, sampling the first site, passing this outcome through the network and sampling the next site, etc. This results in $N$ forward passes in order to generate one exact POVM sample, at the benefit of circumventing Markov chains.

In addition to exact normalization and sampling, we empirically find this network architecture to be substantially easier to train, and the training algorithm to converge for a wider variety of states. There is no simple generalization of this approach to higher dimensional systems, however approaches like \cite{Sharir2019} have been proposed.

The hyperparameter configurations used for producing the data shown in each figure are summarized in Tab.~\ref{tab:hyperparams}.

All training runs throughout this project took less than 2000 epochs to complete while all MLE runs required less than 100 iterations. This shows that there is no prohibitive issue in training-complexity for the states tested. One training epoch using the NN takes a time of $\mathcal{O}(\text{num. Parameters}) = \mathcal{O}(N^3)$ \cite{kingma_adam_2017}, while one iteration in the MLE takes a time of $\mathcal{O}(4^N)$ \cite{lvovsky_iterative_2004}.

\begin{table*}[]
    \centering
    \def\arraystretch{1.75}
    \begin{tabular}{l|c|c|c|c|c}
    Figure & Network architecture & Network boundary conditions & Num. Layers $L$ & Kernel size $K$ & Num. Parameters\\\hline
   ~\ref{fig:ising1d} - Small 1D TFIM & CNN & Circular & $\left[\sqrt{N}\right]$ & $\left\lceil\sqrt{N}\right\rceil + 1$ & 73 - 673\\
   ~\ref{fig:ising2d} - Larger 2D TFIM & CNN & Circular & 2 & 2 & 1329\\
   ~\ref{fig:ion_chain} - Ion Chain & ARCNN & Open & 3 & 6 & 3572\\
   ~\ref{fig:steadyStates} - Steady States & CNN & Open & 2 & 3 & 3169\\
    \end{tabular}
    \caption{Network architectures used for each figure. In all cases we used a tanh-activation function, a learning rate of $10^{-3}$, $N$ features per layer for $N$ particles, and an exponential output function. $[x]$ denotes the nearest integer function and $\lceil x \rceil$ the ceiling function.}
    \label{tab:hyperparams}
\end{table*}

\subsection{Self-averaging can reduce network advantage} 
\begin{figure}[h]
    \centering
    \includegraphics{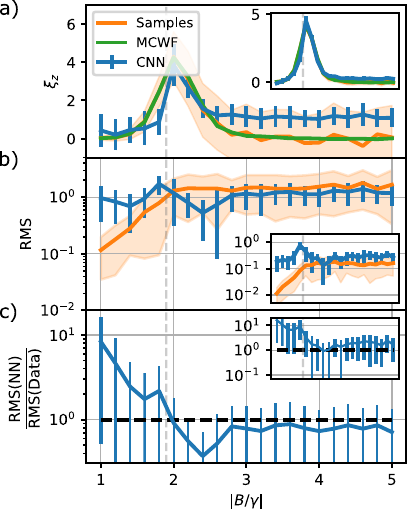}
    \caption{\textbf{Tomography of steady states: demonstration of a self-averaging effect.} Dissipative phase transition of a 4x4 TFIM with spontaneous emission at $J=-1.25\gamma$ with open boundaries. $N_s = 1k$ (insets: $100k$). Gray vertical line: Point where $\uparrow$ and $\downarrow$ have been exchanged in POVM. a) Observables according to Eq.~\eqref{eq:corrLen} summed over \emph{entire} system. b) Residual RMS error. c) Residual network RMS normalized to dataset RMS. \errorbarstatement} 
    \label{fig:steadyStates_full_system}
\end{figure}

As explained in the main text, the effect of a bias, that the network might introduce for a given observable may be amplified when one is interested in observables which contain sums over many independent observables. In this scenario the network-bias inhibits any positive influence of possible self averaging effects. We demonstrate this, by computing the correlation length~\eqref{eq:corrLen} by summing over all pairs of lattice sites, as opposed to only summing over a diagonal as in Fig.~\ref{fig:steadyStates}. The result is shown in Fig.~\ref{fig:steadyStates_full_system}. The sampled observables self-average, resulting in a smaller statistical error at fixed sample size, while the network systematically overestimates the correlation length for fields greater than the critical field. 

\bibliography{references}

\section{Acknowledgments} We thank J.\ Carrasquilla and M.\ Schmitt for discussions.
Monte-Carlo wave function trajectories were computed using the \textsc{Qutip} library \cite{Johansson2013}.
The \textsc{JAX} library \cite{Bradbury2018} as well as the \textsc{Flax} framework \cite{flax2020github} were used to build the neural network models and training algorithms.
This work is supported by the Deutsche Forschungsgemeinschaft (DFG, German Research Foundation) under Germany’s Excellence Strategy EXC2181/1-390900948 (the Heidelberg STRUCTURES Excellence Cluster) and within the Collaborative Research Center SFB1225 (ISOQUANT). This work was partially financed by the Baden-Württemberg Stiftung gGmbH. The authors acknowledge support by the state of Baden-Württemberg through bwHPC
and the German Research Foundation (DFG) through grant no INST 40/575-1 FUGG (JUSTUS 2 cluster). The authors gratefully acknowledge the Gauss Centre for Supercomputing e.V. (www.gauss-centre.eu) for funding this project by providing computing time through the John von Neumann Institute for Computing (NIC) on the GCS Supercomputer JUWELS \cite{JUWELS} at Jülich Supercomputing Centre (JSC).

\section{Data availability}
The code developed for this project for generating test data is available at \href{https://gitlab.com/ann-povm/qst-code}{gitlab.com/ann-povm/qst-code}.

\section{Code availability}
The code developed for this project for performing the NN-QST benchmarks is available at \href{https://gitlab.com/ann-povm/qst-code}{gitlab.com/ann-povm/qst-code}.

\section{Author contributions}
T.S. developed the code base, M.R. performed MCWF simulations. All authors contributed equally to analysing the data and writing of the manuscript. M.G. supervised the project.

\section{Competing interests}
The authors declare no competing interests.

\end{document}